\documentclass[12pt]{article}
\usepackage{epsf}
\usepackage{graphicx}

\begin{document}
\title
{A non-geometrical approach to quantum gravity}
\author
{Michael A. Ivanov \\
Physics Dept.,\\
Belarus State University of Informatics and Radioelectronics, \\
6 P. Brovka Street,  BY 220027, Minsk, Republic of Belarus.\\
E-mail: michai@mail.by.}

\maketitle

\begin{abstract}
Some results of author's work in a non-geometrical approach to
quantum gravity are reviewed here, among them: a quantum mechanism
of classical gravity giving a possibility to compute the Newton
constant; asymptotic freedom at short distances; interaction of
photons with the graviton background leading to the important
cosmological consequences; the time delay of photons due to
interactions with gravitons; deceleration of massive bodies in the
graviton background which may be connected with the Pioneer
anomaly and with the problem of dark matter.
\end{abstract}

\section[1]{Introduction }
Attempts to construct a quantum model of gravity starting from the
geometrical description of classical case in general relativity
may be characterized as very poor in its possible consequences.
The classical limit has not any clear physical mechanism of
formation of the metric, and it is similar to the Newtonian
version where the law of gravity is postulated. There are a few
facts which may be considered as contradicting to the mainstream
in this field of physics: the Pioneer anomaly \cite{1,2}, the
discovery of quantum states of ultra-cold neutrons in the Earth's
gravitational field with very low energies of levels \cite{104};
perhaps, we should include in the list the problem of dark matter
in galaxies. Many people are searching for dark energy, an
existence of which has been claimed on a basis of the standard
cosmological model; but the initial cause for this conclusion may
be put in our list, too.
\par
I would like to review here some results of my work in a
non-geometrical approach to quantum gravity, which is based on the
assumptions that: 1) gravitons are super-strong interacting
particles and 2) the low-temperature graviton background exists.
This model of low-energy quantum gravity has many interesting
consequences, and the one may pave the alternative way to the
future theory. To be shorter, I use here notations of my cited
works.

\section[2]{A quantum mechanism of classical gravity }
It was shown by the author \cite{6,500} that screening the
background of super-strong interacting gravitons creates for any
pair of bodies both attractive and repulsive forces due to
pressure of gravitons. For single gravitons, these forces are
approximately equal. If single gravitons are pairing, an
attractive force due to pressure of such graviton pairs is twice
exceeding a corresponding repulsive force if graviton pairs are
destructed by collisions with a body. In such the model, the
Newton constant may be computed. The attractive force $F_{1}$ due
to pressure of single gravitons in this model is equal to: $
F_{1}\equiv G_{1} \cdot m_{1}m_{2}/r^{2},$ where the constant
$G_{1}$ is: $G_{1} \equiv {1 / 3} \cdot {D^{2} c(kT)^{6} /
{\pi^{3}\hbar^{3}}} \cdot I_{1},$ with $I_{1}= 5.636 \cdot
10^{-3}.$ By $T=2.7~ K:$ $G_{1} =1215.4 \cdot G,$ that is of three
order greater than the Newton constant, $G.$ If single gravitons
are elastically scattered, they create a repulsive force
$F_{1}^{'}$ which is equal to $F_{1}.$ But for black holes which
absorb any particles and do not re-emit them, we will have
$F_{1}^{'} =0.$ It means that such the objects would attract other
bodies with a force which is proportional to $G_{1}$ but not to
$G,$ i.e. Einstein's equivalence principle would be violated for
them.
\par
In a case of graviton pairing, a force of attraction of two bodies
$F_{2}$ due to pressure of graviton pairs will be equal to:
$F_{2}= {8 / 3} \cdot {D^{2} c(kT)^{6} m_{1}m_{2} /
{\pi^{3}\hbar^{3}r^{2}}}\cdot I_{2},$ where $I_{2} = 2.3184 \cdot
10^{-6}.$ The difference $F$ between attractive and repulsive
forces is twice smaller: $F \equiv F_{2}- F_{2}^{'}={1 / 2}F_{2}
\equiv G_{2}{m_{1}m_{2} / r^{2}},$ where the constant $G_{2}$ is:
$G_{2} \equiv {4 / 3} \cdot {D^{2} c(kT)^{6} / {\pi^{3}\hbar^{3}}}
\cdot I_{2}.$ If one assumes that $G_{2}=G,$ then it gives for the
new constant $D$: $D=0.795 \cdot 10^{-27}{m^{2} / eV^{2}}.$
\par
The inverse square law takes place in the model if the condition
of big distances is fulfilled: $ \sigma (E,<\epsilon>) \ll 4 \pi
r^{2}$ \cite{6,500}. It leads to the necessity of some "atomic
structure" of matter for working the described quantum mechanism;
it is a unique demand for known models of gravity.
\section[3]{Asymptotic freedom at short distances }
Recently, it was shown in \cite{116} that asymptotic freedom
appears at very short distances in this model. In this range, the
screened portion of gravitons tends to the fixed value of $1/2$,
that leads to the very small limit acceleration of the order of
$10^{-13} \ m/s^{2}$ of any screened micro-particle.
\begin{figure}[htb]
\epsfxsize=6.0cm \centerline{\epsfbox{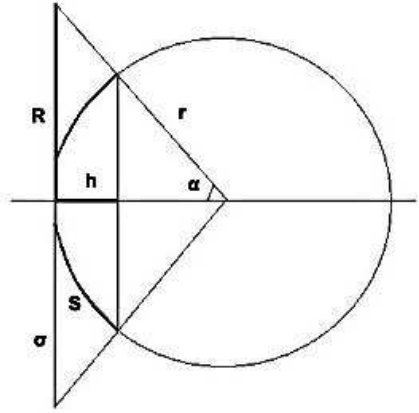}} \caption{To the
computation of the screened portion of gravitons at small
distances: $\sigma$ is the cross-section, $S$ is a square of the
spherical segment of a hight $h$. }
\end{figure}
The ratio $\sigma (E_{2},<\epsilon_{2}>) / 4 \pi r^{2}$ describes
the screened portion of gravitons for a big distance $r$. For
small $r$, let us consider Fig. 1, where $R=(\sigma
(E_{2},<\epsilon_{2}>)/\pi)^{1/2}$, $S$ is the screening area. It
is necessary to replace the ratio $\sigma (E_{2},<\epsilon_{2}>) /
4 \pi r^{2}$ by the following one: $\rho(y) \equiv S / 4 \pi
r^{2}.$
\par
To find the net force of gravitation $F$ at a small distance $r$,
we should replace the factor $\sigma (E_{2},<\epsilon_{2}>) / 4
\pi r^{2}$ in Eq. (31) of \cite{500} with the more exact factor $S
/ 4 \pi r^{2}$. Then we get:
\begin{equation}
F(r)= {4 \over 3} \cdot {D(kT)^{5}E_{1} \over
\pi^{2}\hbar^{3}c^{3}} \cdot g(r),
\end{equation}
where $E_{1}$ is an energy of particle 1, and  $g(r)$ is the
function of $r$:
\begin{equation}
g(r) \equiv \int_{0}^{\infty}{ x^{4}
(1-\exp(-(\exp(2x)-1)^{-1}))(\exp(2x)-1)^{-3} \over
\exp((\exp(2x)-1)^{-1}) \exp((\exp(x)-1)^{-1})} \cdot \rho(y) d x,
\end{equation}
where $y=y(r,x)=r/R(x)$. By $r\rightarrow 0$, this function's
limit for any $E_{2}$ is: $g(r)\rightarrow I_{5}= 4.24656 \cdot
10^{-4}$. For comparison, graphs of the function $g(r)$ are shown
in Fig. 2 for
\begin{figure}[htb]
\epsfxsize=6.0cm \centerline{\epsfbox{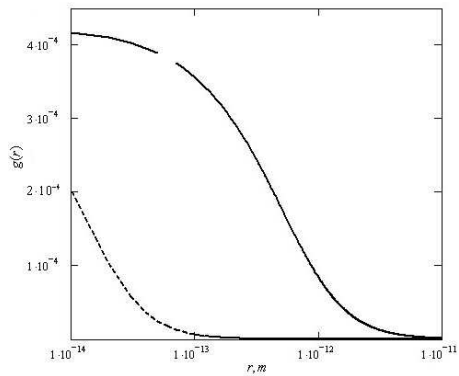}} \caption{Different
transition to the limit value of the function $g(r)$ by
$E_{2}=m_{p}c^{2}$ (solid) and by $E_{2}=m_{e}c^{2}$ (dot).}
\end{figure}
the following different energies: $E_{2}=m_{p}c^{2}$ and
$E_{2}=m_{e}c^{2}$. The functions have the same limit by
$r\rightarrow 0$, but the most interesting thing is their
different transition to this limit when $r$ decreases. The range
of transition for a proton is between $10^{-11} - 10^{-13}$ meter,
while for an electron it is between $10^{-13} - 10^{-15}$ meter.
\par The property of asymptotic freedom leads to the important consequence:
a black hole mass threshold should exist \cite{216,217}.
\section[4]{Interaction of photons with the graviton background }
Due to forehead collisions with gravitons, an energy of any photon
should decrease when it passes through the sea of gravitons. From
another side, none-forehead collisions of photons with gravitons
of the background will lead to an additional relaxation of a
photon flux, caused by transmission of a momentum transversal
component to some photons. It will lead to an additional dimming
of any remote objects, and may be connected with supernova
dimming. Average energy losses of a photon with an energy  $E $ on
a way $dr $ will be equal to: $dE=-aE dr,$ where $a=H/c$. In this
model, $H= {1 / 2\pi}\cdot D \cdot \bar \epsilon \cdot (\sigma
T^{4}),$ where $\bar \epsilon$ is an average graviton energy. As a
result, we have: $E(r)=E_{0} \exp(-ar), $ where $E_{0}$ is an
initial value of energy. Both redshifts and the additional
relaxation of any photonic flux due to non-forehead collisions of
gravitons with photons lead in the model to the following
luminosity distance $D_{L}:$ $D_{L}=a^{-1} \ln(1+z)\cdot
(1+z)^{(1+b)/2} \equiv a^{-1}f_{1}(z),$ where $f_{1}(z)\equiv
\ln(1+z)\cdot (1+z)^{(1+b)/2}$, with the factor $b\simeq 2.137$
for soft radiation. It is easy to find a value of the factor $b$
in another marginal case - for a very hard radiation. Due to very
small ratios of graviton to photon momenta, photon deflection
angles will be small, but collisions will be frequent because the
cross-section of interaction is a bilinear function of graviton
and photon energies in this model. It means that in this limit
case $b \rightarrow 0.$ For an arbitrary source spectrum, a value
of the factor $b$ should be still computed. It is clear that $0
\leq b \leq 2.137,$ and in a general case it should depend on a
rest-frame spectrum and on a redshift. It is important that the
Hubble diagram in the model is a multivalued function of a
redshift: for a given $z,$ $b$ may have different values
\cite{115}.
\par Using only the luminosity distance and a geometrical one as functions of
a redshift in this model, theoretical predictions for
galaxy/quasar number counts may be found \cite{119}. For example,
galaxy number counts as a function of a redshift $z$ may be
characterized with the function $f_{2}(z)$ for which we have in
this model: $f_{2}(z)= {ln^{2}(1+z) / {z^{2}(1+z)}}.$ A graph of
this function is shown in Fig. 3; the typical error bar and data
point are added here from paper \cite{72} by Loh and Spillar.
There is not a visible contradiction with observations. {\it There
is not any free  parameter in the model to fit this curve;} it is
a very rigid case.
\begin{figure}[th]
\epsfxsize=6.0cm \centerline{\epsfbox{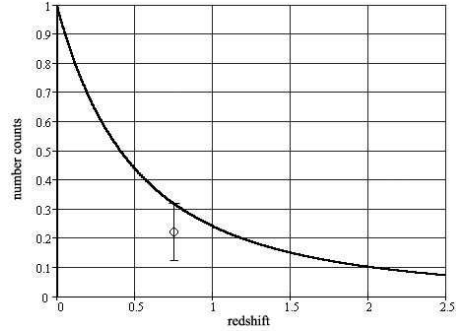}} \caption{Number
counts $f_{2}$ as a function of the redshift in this model. The
typical error bar and data point are taken from paper \cite{72} by
Loh and Spillar.}
\end{figure}
\section[5]{Time delay of photons due to interactions with gravitons }
To compute the average time delay of photons, it is necessary  to
find a number of collisions on a small way $dr$ and to evaluate a
delay due to one act of interaction. Let us consider at first the
background of single gravitons. Given the expression for $H$ in
the model, we can write for the number of collisions with
gravitons having an energy $\epsilon =\hbar \omega$:
\begin{equation}
dN(\epsilon)= {|dE(\epsilon)| \over \epsilon} = E(r)\cdot {dr
\over c} {1 \over 2\pi}  D f(\omega,T)d \omega,
\end{equation}
where $f(\omega,T)$ is described by the Plank formula. In the
forehead collision, a photon loses the momentum $\epsilon/c$ and
obtains the energy $\epsilon$; it means that for a virtual photon
we will have:
\begin{equation}
{v\over c}={{E- \epsilon}\over {E+ \epsilon}}; \ 1- {v\over c}={
2\epsilon \over {E+ \epsilon}}; \ \ 1- {v^2\over c^2}={ 4\epsilon
E \over (E+ \epsilon)^2}.
\end{equation}
The uncertainty of energy for a virtual photon $\Delta E=2
\epsilon$. If we evaluate the lifetime of a virtual photon on a
basis of the uncertainties relation: $\Delta E \cdot \Delta \tau
\geq \hbar/2 $, we get $\Delta \tau \geq \hbar/4\epsilon$. In the
time $\Delta \tau$, the time delay $\Delta t$ will be equal to:
\begin{equation}
\Delta t(\epsilon)=\Delta \tau (1- {v\over c})\geq \hbar/2 \cdot
{1 \over {E+ \epsilon} }.
\end{equation}
The full time delay due to gravitons with an energy $\epsilon$ is:
$dt(\epsilon)=\Delta t(\epsilon) dN(\epsilon)$. Taking into
account all frequencies, we find the full time delay on the way
$dr$:
\begin{equation}
dt\geq  \int_{0}^{\infty} {\hbar \over 2}{E  \over
{E+\epsilon}}\cdot {dr \over c} {1 \over 2\pi}D f(\omega,T)d
\omega.
\end{equation}
The full time delay on the way $dr$ will be maximal for $E
\rightarrow \infty, $ and it is easy to evaluate the one:
\begin{equation}
dt_{\infty} \geq  {\hbar \over 4\pi}{dr \over c}\cdot D \sigma
T^4.
\end{equation}
On the way $r$ the time delay is:
\begin{equation}
t_{\infty}(r) \geq  {\hbar \over 4\pi}{r \over c}\cdot D \sigma
T^4.
\end{equation}
In this model: $r(z)=c/H\cdot \ln(1+z)$; let us introduce a
constant: $\rho \equiv {\hbar / 4\pi}\cdot D \sigma T^4/H = 37.2
\cdot 10^{-12} s$, then
\begin{equation}
t_{\infty}(z) \geq \rho \ln(1+z).
\end{equation}
We see that for $z\simeq 2 $ the maximal time delay is equal to
$\sim 40 \ ps$, i.e. the one is negligible. If we take into
account graviton pairing, the estimate of delay becomes smaller.
\par If we consider another possibility of lifetime estimation, for example,
$\Delta \tau_{0}=const$, where $\Delta \tau_{0}$ is the proper
lifetime of a virtual photon (it should be considered as a new
parameter of the model), taking into account that now:
\begin{equation}
\Delta \tau =\Delta \tau_{0}/(1- {v^2\over c^2})^{1/2},
\end{equation}
we shall get in the same manner (my paper about the time delay is
now in progress):
\begin{equation}
t(z) =\Delta \tau_{0} \sqrt{E_{0}/\epsilon_{0}} \cdot {{\sqrt{1+z}
-1} \over \sqrt{1+z}},
\end{equation}
where $E_{0}$ is an initial photon energy, $\epsilon_{0}$ is a new
constant: $\epsilon_{0}=2.391 \cdot 10^{-4} \ eV$.
\par In this case, the time delay of photons with different
initial energies $E_{01}$ and $E_{02}$ will be proportional to the
difference $\sqrt{E_{01}}-\sqrt{E_{02}}$, and more energetic
photons should arrive later, also as in the first case. It is
still necessary to calculate the dispersion of the delay. To find
$ \Delta \tau_{0}$, we must compare the computed value of time
delay with future observations. Recently, an analysis of
time-resolved emissions from the gamma-ray burst GRB 081126
\cite{372} showed that the optical peak occurred $(8.4 \pm 3.9) \
s$ {\it later} than the second gamma peak; perhaps, it means that
this delay is connected with the mechanism of burst.
\section[6]{Deceleration of massive bodies in the graviton background}
The observed Pioneer anomaly \cite{1,2} has the following main
features: 1) in the range 5 - 15 AU from the Sun it is observed an
anomalous sunward acceleration with the rising modulus which gets
its maximum value; 2) for greater distances, this maximum sunward
acceleration remains almost constant for both Pioneers; 3) it is
observed an unmodeled annual periodic term in residuals for
Pioneer 10 \cite{311} which is obviously connected with the motion
of the Earth. In a frame of this model, a universal character of
gravitational interaction should lead to energy losses of any
massive body due to forehead collisions with gravitons, so the
body acceleration $w \equiv dv/dt$ by a non-zero velocity $v$ is
equal to: $w = - ac^{2}(1-v^{2}/c^{2}).$  For small velocities: $w
\simeq - Hc.$ If the Hubble constant $H$ is equal to its
theoretical estimate in this approach $2.14 \cdot 10^{-18}
s^{-1},$ a modulus of the acceleration will be equal to $|w|
\simeq Hc = 6.419 \cdot 10^{-10} \ m/s^{2},$ that is of the same
order of magnitude as  a value of the observed additional
acceleration $(8.74 \pm 1.33) \cdot 10^{-10} m/s^2$ for NASA
probes. The acceleration $w$ is directed against a body velocity
in the frame of reference in which the graviton background is
isotropic. This acceleration will have different directions by
motion of a body on a closed orbit. The observed value of
anomalous acceleration of Pioneer 10 should represent the vector
difference of the two accelerations \cite{6}: an acceleration of
Pioneer 10 relative to the graviton background, and  an
acceleration of the Earth relative to the background. Perhaps,
namely the last one is displayed as an annual periodic term in the
residuals of Pioneer 10 \cite{311}. An observed value of the
projection of the probe's acceleration on the sunward direction
$w_{s}$ should depend on accelerations of the probe, the Earth and
the Sun relative to the graviton background. If the Sun moves
relative to the background slowly enough, then anomalous
accelerations of the Earth and the probe would be directed almost
against their velocities in the heliocentric frame, and in this
case:  $w_{s}=-w\cdot cos\alpha$, where $\alpha$ is an angle
between a radius-vector of the probe and its velocity in the
frame. By the very elongate orbits of the both Pioneers, it would
explain the second (and main) peculiarity. For example, for
Pioneer 10 at the distance 67 AU from the Sun one has
$\sin\alpha\approx0.11$, i.e. $cos\alpha\approx0.994.$ If for big
distances from the Sun we use the conservation laws of energy and
angular momentum in the field {\it of the Sun only}, then in the
range 6.7 - 67 AU a value of $cos\alpha$ changes from 0.942 to
0.994, i.e. approximately on 5 per cent only. Due to this fact, a
projection of the probe's acceleration on the sunward direction
would be almost constant \cite{300}.
\par
As Toth and Turyshev report \cite{3}, they intend to carry out an
analysis of newly recovered data received from Pioneers, with
these data are now available for Pioneer 11 for distances 1.01 -
41.7 AU. If the serious problem of taking into account the solar
radiation pressure at small distances is precisely solved
(modeled) \cite{4}, then this range will be very lucky to confront
the expression $w_{s}=-w\cdot cos\alpha$ of the considered model
with observations for small distances when Pioneer 11 executed its
planetary encounters with Jupiter and Saturn. In this period, a
value of $cos\alpha$ was changed in the non-trivial manner. For
example, when the spacecraft went to Saturn, $cos\alpha$ was {\it
negative} during some time. If this model is true, the anomaly in
this small period should have {\it the opposite sign}. It would be
the best of all to compare the two functions of the probe's proper
time: the projection of anomalous acceleration of Pioneer 11 and
$cos\alpha$ for it. These functions should be very similar to each
other if my conjecture is true. At present, a new mission to test
the anomaly is planned \cite{44}. It is seen from this
consideration that it would be desirable to have a closed orbit
for this future probe, or the one with two elongate branches where
the probe moves off the Sun and towards it.
\par This deceleration of massive bodies by the graviton
background may lead to an additional {\it relative} acceleration
of bodies in a closed system. For example, when a galaxy moves
through the background, a deceleration of its center will be
constant, but for orbiting it stars the same deceleration will
change its sign. The kinetic energy of stars should increase with
time in the rest frame of the center. Perhaps, namely the fact
obeys successes of MOND by M. Milgrom in explanation of flat
rotation curves of galaxies \cite{99} (and its failure for
clusters of galaxies). In MOND, when a body acceleration gets the
threshold value of $\sim Hc,$ one introduces by hand the growth of
interaction; but namely this value characterizes the Pioneer
anomaly in this model.
\section[7]{Conclusion}
I hope that this model may help us to see and to realize some
fresh ideas in the very old area. The coincidence of the magnitude
of the anomalous deceleration of Pioneer 10/11 with the product
value of $Hc$, and an appearance of the same quantity in the MOND
cannot be due to a chance. Many consequences of the model have an
impact on our understanding of cosmological problems, and these
very close ties between micro and macro cosmoses are very exiting.


\begin{thebibliography}{References                        }
\bibitem{1}
Anderson, J.D. et al. {\it Phys. Rev. Lett.} 1998, {\it 81,} 2858.
\bibitem{2}
Anderson, J.D. et al. {\it Phys. Rev.} 2002, {\it D65,} 082004;
[gr-qc/0104064 v4].
\bibitem{104}
Nesvizhevsky, V.V. et al. {\it Nature} 2002, {\it 415,} 297.
\bibitem{6}
Ivanov, M.A. [gr-qc/0207006].
\bibitem{500}
Ivanov, M.A. In the book "Focus on Quantum Gravity Research", Ed.
D.C. Moore, Nova Science, NY - 2006 - pp. 89-120;
[hep-th/0506189], [http://ivanovma.narod.ru/nova04.pdf].
\bibitem{116}
Ivanov, M.A. {\it J. Grav. Phys.}, 2008, V.2, No.2, pp. 26-31;
[arXiv:0801.1973v1 [hep-th]].
\bibitem{216}
Ivanov, M.A. [arXiv:0901.0510v1 [physics.gen-ph]].
\bibitem{217}
Ivanov, M.A. Gravitational asymptotic freedom and matter filling
of black holes. Contribution to PIRT-09, Moscow, 6-9 July 2009.
\bibitem{115}
Ivanov, M.A. [astro-ph/0609518v4].
\bibitem{119}
Ivanov, M.A. [astro-ph/0606223].
\bibitem{72}
Loh, E.D. and Spillar, E.J. {\it ApJ} 1986, {\it 307,} L1.
\bibitem{372}
Klotz, A. et al. ApJ Lett. 697 (2009) L18; [arXiv:0904.4786v1
[astro-ph.CO]].
\bibitem{311}
Turyshev, S.G. et al. [gr-qc/9903024 v2].
\bibitem{300}
Ivanov, M.A. [arXiv:0711.0450v2 [physics.gen-ph]].
\bibitem{3}
Toth, V.T. and Turyshev, S.G. [arXiv:0710.2656v1 [gr-qc]].
\bibitem{4}
Anderson, J.D. and Nieto, M.M. {\it Contemp. Phys.} 2007, {\it
48}, No. 1. 41-54; [arXiv:0709.3866v1 [gr-qc]].
\bibitem{44}
Dittus, H. et al. [gr-qc/0506139].
\bibitem{99}
Milgrom, M. [astro-ph/9810302].

\end{thebibliography}
\end{document}